\documentclass{emulateapj}
\usepackage{natbib}
\usepackage{amsfonts,amsmath,amssymb}
\usepackage{graphicx}
\usepackage{color}
\usepackage[backref,breaklinks,colorlinks,citecolor=blue]{hyperref}
\usepackage{url}
\usepackage{acronym}

\newacro{HST}{Hubble Space Telescope}
\newacro{RGB}{Red Giant Branch}
\newacro{AGB}{Asymptotic Giant Branch}
\newacro{CMD}{Color Magnitude Diagram}
\newacro{ACS}{Advanced Camera for Surveys}
\newacro{WFC}{Wide Field Channel}
\newacro{TRGB}{Tip of the Red Giant Branch}
\newacro{AST}{Artificial Star Test}
\newacro{SDSS}{Sloan Digital Sky Survey}
\newacro{PSF}{Point Spread Function}
\newacro{LG}{Local Group}
\newacro{MCMC}{Markov Chain Monte Carlo}
\newacro{SFH}{star formation history}
\newacroplural{SFH}[SFHs]{star formation histories}

\providecommand{\farcsec}{.\!\!^{\prime\prime}}%  % fractional arcsecond symbol: 0.''0
\providecommand{\fras}{.\!\!^{s}}%  % fractional seconds of ra symbol: 0.s0
\newcommand{\unifdistsymb}{{\rm U}}

\begin{document}

\title{HST Imaging of the Local Volume Dwarf Galaxies Pisces A\&B: Prototypes for Local Group Dwarfs
}
\shorttitle{ACS Imaging of Pisc Dwarfs}


\author{Erik J.\ Tollerud\altaffilmark{1,2,7},
        Marla C.\ Geha\altaffilmark{2},
        Jana Grcevich\altaffilmark{3},
        Mary E.\ Putman\altaffilmark{4},
        Daniel R.\ Weisz\altaffilmark{5,7},
        Andrew E.\ Dolphin\altaffilmark{6}
       }

\altaffiltext{1}{Space Telescope Science Institute, 3700 San Martin Dr, Baltimore, MD 21218, USA; etollerud@stsci.edu}
\altaffiltext{2}{Astronomy Department, Yale University, P.O. Box 208101, New Haven, CT 06510, USA}
\altaffiltext{3}{Department of Astrophysics, American Museum of Natural History, Central Park West at 79th St., New York, NY 10024,
USA}
\altaffiltext{4}{Department of Astronomy, Columbia University, New York, NY 10027, USA}
\altaffiltext{5}{Department of Astronomy, University of Washington, Box 351580, Seattle, WA 98195, USA}
\altaffiltext{6}{Raytheon Company, 1151 East Hermans Road, Tucson, AZ 85756, USA}
\altaffiltext{7}{Hubble Fellow}

\begin{abstract}We present observations of the Pisces A and B galaxies with the \acl{ACS} on the \acl{HST}.
Photometry from these images clearly resolve a \acl{RGB} for both objects, demonstrating that they are nearby dwarf galaxies.
We describe a Bayesian inferential approach to determining the distance to these galaxies using the magnitude of the \acl{TRGB}, and then apply this approach to these galaxies.
%We also provide the full probability distributions for parameters derived using this approach.
This reveals the distance to these galaxies as $5.64^{+0.13}_{-0.15} \, {\rm Mpc}$ and $8.89^{+0.75}_{-0.85} \, {\rm Mpc}$ for Pisces A and B, respectively, placing both within the Local Volume but not the Local Group.
We  estimate the \aclp{SFH} of these galaxies, which suggests that they have recently undergone an increase in their star formation rates.
Together these yield luminosities for Pisces A and B of $M_V=-11.57^{+0.06}_{-0.05}$  and $-12.9 \pm 0.2$, respectively, and estimated stellar masses of $\log(M_*/M_{\odot})= 7.0^{+0.4}_{-1.7}$ and $7.5^{+0.3}_{-1.8}$.
We further show that these galaxies are likely at the boundary between nearby voids and higher-density filamentary structure.
This suggests that they are entering a higher-density region from voids, where they would have experienced delayed evolution, consistent with their recent increased star formation rates.
If this is indeed the case, they are useful for study as proxies of the galaxies that later evolved into typical \acl{LG} satellite galaxies.
\end{abstract}

\section{Introduction}

Faint dwarf galaxies provide constraints on dark matter and cosmology \citep[e.g.][]{kl99ms, krav10satrev, BKBK11} as well as tests of the complex  physics behind galaxy formation \citep[e.g.,][]{dekelandsilk, geha06, weisz11b, brooks14}.
A major reason for this is that the nearest dwarfs can be studied in resolved stars because they are both faint and nearby enough that crowding issues can be overcome for a relatively large sample.
By characterizing their resolved stellar populations, it becomes possible both to obtain present-day structural parameters for these galaxies \emph{and} characterize their \acp{SFH}, providing constraints on their past \citep[e.g.,][]{lgtime}.
While this is powerful, it does not provide a way to determine their past structural properties or gas content.
Furthermore, such faint galaxies are primarily observable only in the very nearby universe, where the bright galaxies of the \ac{LG} seem to quench the dwarfs \citep[e.g.,][]{einasto74, fill15, wetzel15}.

This suggests that searching for faint dwarf galaxies in other environments might provide further clues to dwarf galaxy formation and evolution.
Recently such objects have been identified from compact 21 cm \ion{H}{1} clumps at velocities consistent with nearby dwarf galaxies, which are followed-up on via optical imaging to confirm the presence of dwarf galaxies.
The Leo P dwarf was found using this approach using data from ALFALFA \citep{leop}, and compact clouds in the GALFA-HI Survey led to detection of the Pisces A and B galaxies \citep{T15} and two others \citep{Sand15}.
These GALFA-HI galaxies overlap in properties with some of the faintest already-known dwarfs \citep[e.g., the sample from][]{SHIELD}, but without firm distance measures it is difficult to place them in a wider galaxy formation context.
Hence, here we present \ac{HST} \ac{ACS}/\ac{WFC} imaging of Pisces A and B.
The unmatched resolution available from \ac{HST} allows resolving their stellar populations, identifying an \ac{RGB}, and thereby obtaining distances to these galaxies.

This paper is organized as follows: in Section \ref{sec:obs}, we describe the \ac{HST} observations of the Pisces A and B dwarf galaxies. In Section \ref{sec:dist}, we describe our method for determining \ac{RGB} distances, as well as the resulting distance estimates.  In Section \ref{sec:sparams}, we discuss the structural parameters of the galaxies.  In Section \ref{sec:sfhs}, we fit the \acp{CMD} to determine \acp{SFH} of these dwarfs.  In Section \ref{sec:disc}, we provide context for these galaxies and in Section \ref{sec:conc} we conclude. To aid reproducibility, the analysis software used for this paper is available
at \url{https://github.com/eteq/piscdwarfs_hst}; full Mar kov Chain Monte Carlo (MCMC) chains are also made available as an online data set at \url{http://dx.doi.org/10.5281/zenodo.51375}.

In Table \ref{tab:res} we provide an overview of the key properties of Pisces A and B.  While we describe the methods used to derive these values in the remainder of this paper, we provide the table here for easy reference. When describing uncertainties of the quantities in this table and other parts of this paper, we will use the median and 84th/16th percentile (i.e., $1\sigma$ confidence region).  However, it is important to recognize that some of these quantities are non-Gaussian.  Hence, we provide samples from these distributions in Table \ref{tab:quantities} to allow better modeling of any of these quantities in other contexts or follow-on work.

\begin{table}[htbp]

\begin{center}
\caption{Key properties of Pisces A and B.}
\label{tab:res}

\begin{tabular}{l  l  c  c }

  \hline
  \hline
 & & Pisces A &  Pisces B   \\
 \hline
  (1) & R.A. (J2000)                                     & $00^{\rm h}14^{\rm m}46\fras0 $ & $01^{\rm h}19^{\rm m}11\fras7 $ \\
  (2) & Dec (J2000)                                      & $+10^{\circ}48^{\prime}47\farcsec01$ & $+11^{\circ}07^{\prime}18\farcsec22$ \\
  (3) &  $b$ ($^\circ$)                                    & -51.03 & -51.16 \\
  (4) &  $l$ ($^\circ$)                                    & 108.52 & 133.83 \\
  (5) &  Distance (Mpc)                                   & $5.64^{+0.15}_{-0.13}$ & $8.89^{+0.75}_{-0.85}$ $\dagger$ \\
  (6) & Distance modulus (mag)                            & $28.76^{+0.05}_{-0.06}$ & $29.75^{+0.19}_{-0.20}$ $\dagger$ \\
  (7) &  $\mu_{\rm F814W}$ (mag)                       & $24.62 \pm 0.05$ & $25.61^{+0.19}_{-0.18}$ $\dagger$ \\
  (8) &  $F606W_0$ (mag)                                    & $-11.67^{+0.06}_{-0.05}$ & $-12.9 \pm 0.2$ $\dagger$ \\
  (9) &  $F814W_0$ (mag)                                    & $-12.31 \pm 0.06$ & $-13.4 \pm 0.2$ $\dagger$ \\
  (10) &  $M_V$ (mag)                                    & $-11.57^{+0.06}_{-0.05}$ & $-12.9 \pm 0.2$ $\dagger$ \\
  (11) &  $V-I$ (mag)                                    & $0.78 \pm 0.01$ & $0.57 \pm 0.01$ \\
  (12) &  $R_{\rm eff, major}$ ($^{\prime\prime}$)                                    & $9.1 \pm 0.1$ & $10.37 \pm 0.03$ \\
  (13) &  $r_{\rm eff}$ (pc)                                    & $145^{+5}_{-6}$ & $323^{+27}_{-30}$ $\dagger$ \\
  (14) &  $n$                                     & $0.47^{+0.02}_{-0.01}$ & $0.64 \pm 0.01$ \\
  (15) &  $e$                                    & $0.34 \pm 0.01$ & $0.52 \pm 0.01$ \\
  (16) &  $\theta$ ($^\circ$ E of N)                        & $136 \pm 1.3$ & $139 \pm 0.1$ \\
  (17) & $\log{(M_{*,SFH}/M_{\odot})}$              & $7.0^{+0.4}_{-1.7}$ & $7.5^{+0.3}_{-1.8}$ \\
  (18) & $M_{\rm HI}$ ($10^6 M_{\odot}$)                   & $8.9 \pm 0.8$ & $30^{+6}_{-7}$ $\dagger$ \\
  (19) & $v_{\rm helio, HI}$ (${\rm km \; s}^{-1}$)      & $236 \pm 0.5$ & $615 \pm 1$ \\
  (20) & $W50_{\rm HI}$ (${\rm km \; s}^{-1}$)           & $22.5 \pm 1.3$ & $43 \pm 3$ \\
  (21) & $v_{\rm helio, opt}$ (${\rm km \; s}^{-1}$)     & $240 \pm 34$ & $607 \pm 35$ \\
  \hline

\end{tabular}
\end{center}

Row (1)-(2) On-sky equatorial coordinates. (3)-(4) Galactic latitude/longitude. (5)-(6) Distance/distance modulus from \ac{TRGB} as described in \S \ref{sec:dist}. (7) F814W magnitude of the \ac{TRGB}. (8)-(9) Absolute F606W and F814W magnitudes (\S \ref{sec:sparams}). (10)-(11) VI magnitude and color, transformed from the \ac{HST} bands following the prescription of \citet{acstrans05}, \S 8.3. (12) On-sky major axis half-light radius from F606W imaging (\S \ref{sec:sparams}). (13) Physical half-light radius (circularized) from F606W imaging. (14) S{\'e}rsic index from F606W imaging (\S \ref{sec:sparams}). (15) Ellipticity ($1-b/a$) from F606W imaging (\S \ref{sec:sparams}). (16) Position angle of major axis from F606W imaging (\S \ref{sec:sparams}). (17) Total stellar mass as inferred from the \ac{CMD} from fitting the \ac{SFH} following the procedure described in \S \ref{sec:sfhs}. (18) \ion{H}{1} gas mass from GALFA-HI \citep{Peek11galfadr1} as described in \citet{T15}, assuming the distance inferred in \S \ref{sec:dist}. (19) systemic velocity of \ion{H}{1} gas. (20) W50 of \ion{H}{1} gas. (21) optical velocity, from the H$\alpha$ emission lines reported in \citet{T15}.

$\dagger$: A quantity that has a significantly non-Gaussian distribution.  Such quantities are better represented using the samples from the distributions given in Table \ref{tab:quantities}.

\end{table}

\section{ACS Observations, Reductions, and CMDs}
\label{sec:obs}

\begin{figure}[]
\begin{center}
\includegraphics[width=1\columnwidth]{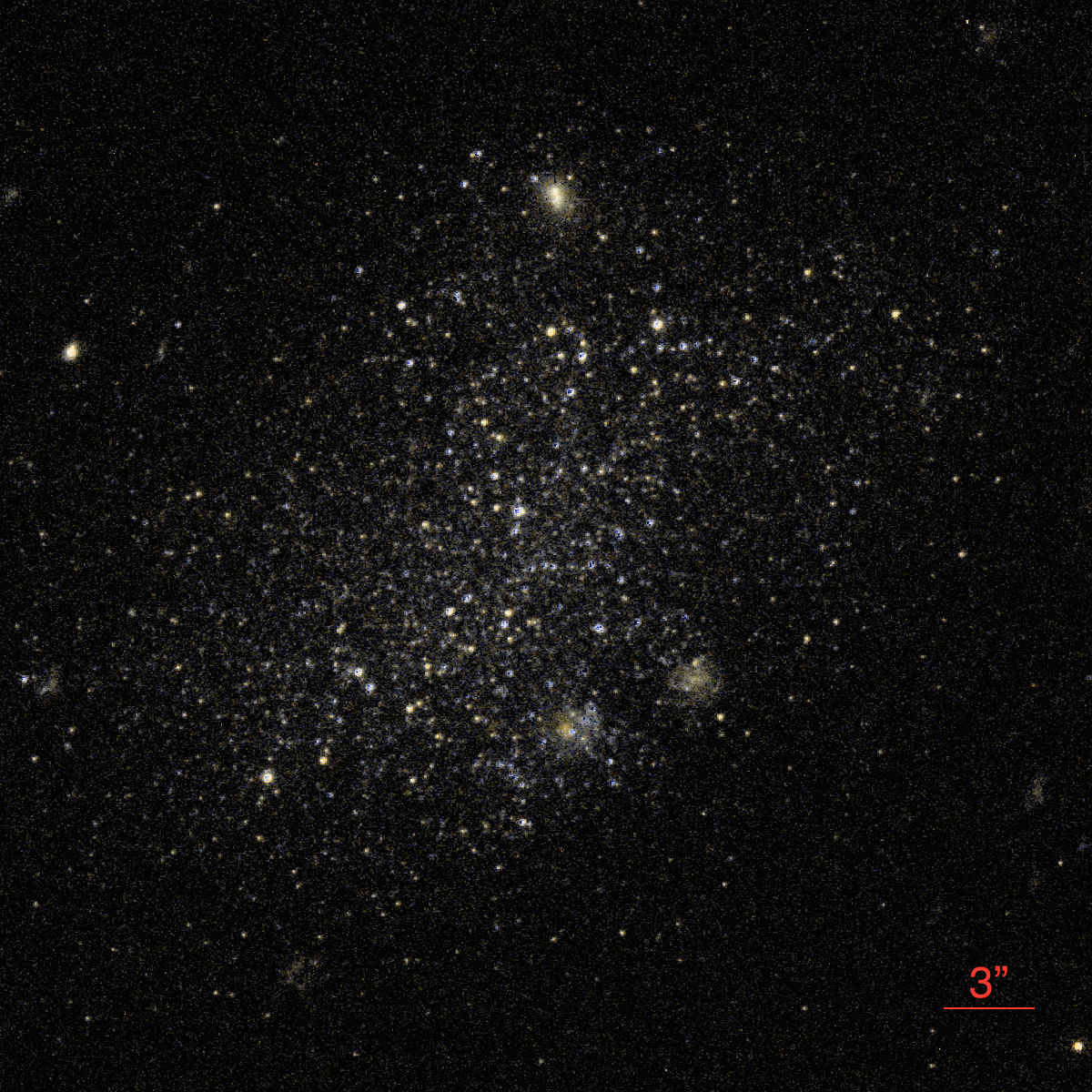}
\caption{ \protect\label{fig:piscAimg}
Color composite of F606W/F814W \ac{HST} \ac{ACS} imaging of the dwarf galaxy Pisces A.  This color composite approximates colors the human eye would see by fitting a blackbody to the per-pixel F606W/F814W flux and assigns RGB colors by convolving that blackbody with response functions of the human eye's cones \citep{Stockman_1993}. The pixel scale is $0.03^{\prime\prime}/{\rm pixel}$.}
\end{center}
\end{figure}

\begin{figure}[]
\begin{center}
\includegraphics[width=1\columnwidth]{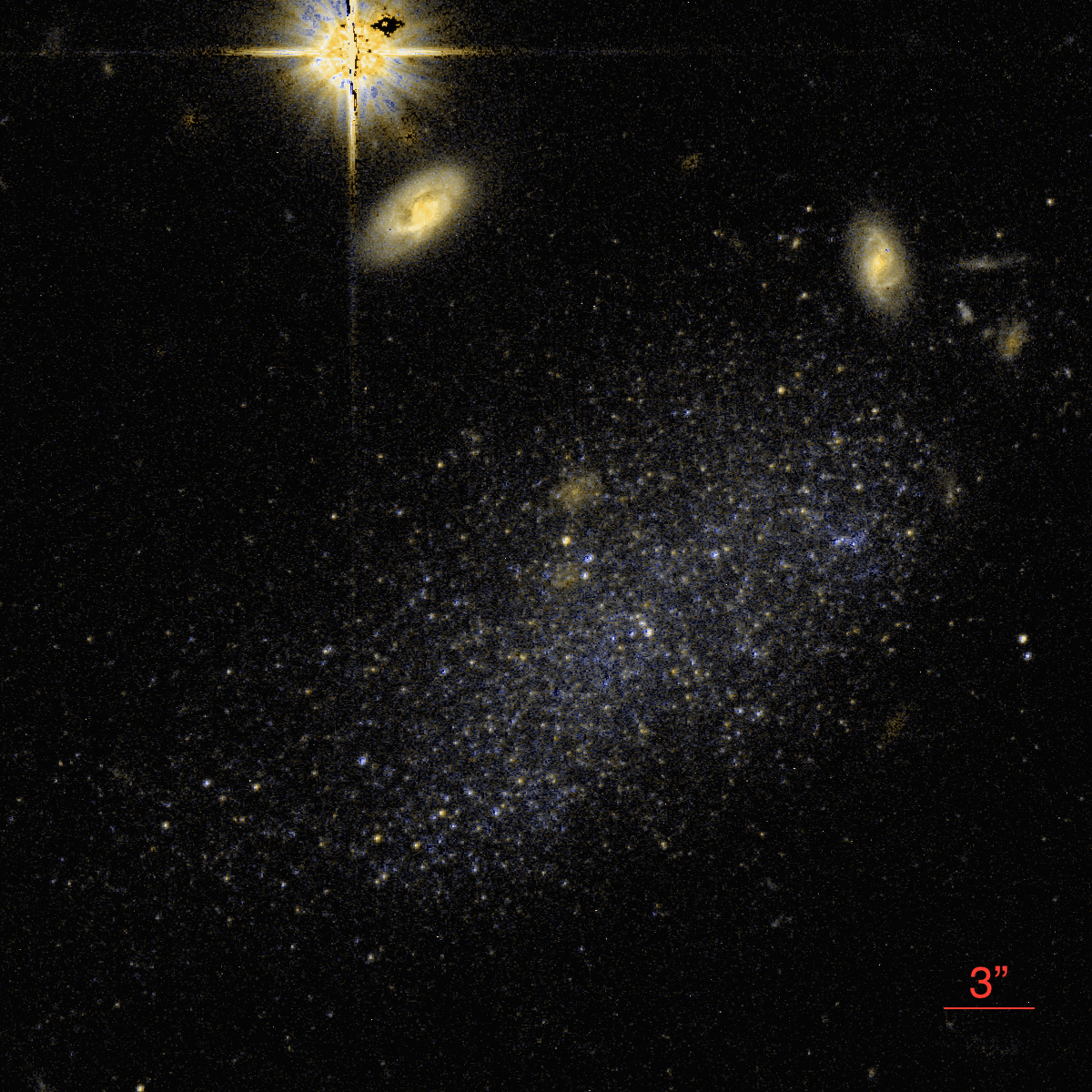}
\caption{ \protect\label{fig:piscBimg}
Same as Figure \ref{fig:piscAimg}, but for Pisces B.}
\end{center}
\end{figure}

The observations of Pisces A and B were taken as part of \ac{HST} program GO-13745 (Cycle 22, PI Tollerud, executed Oct 30 to Nov 8 2014). Pisces A and B were observed for a single orbit each with \ac{ACS}/\ac{WFC}, using the F606W and F814W filters.  For Pisces A the total exposure times were 1092 and 1040 seconds for F606W and F814W, respectively, and for Pisces B, 1072 and 1020 seconds. Color composites of these observations are shown in Figures \ref{fig:piscAimg} and \ref{fig:piscBimg}.

For both targets, we dithered two exposures per filter, to fill in the chip gap (although it does not pass through the central body of either dwarf) and for cosmic ray rejection.  Each image was processed through the standard CALACS pipeline (v8.3.0) and corrected for charge transfer efficiency degradation using v3.3 of the \citet{ctecorr} pixel-based correction scheme.  When working with combined images (Figures \ref{fig:piscAimg} and \ref{fig:piscBimg} and in \S \ref{sec:sparams}), we use {\it Drizzlepac/AstroDrizzle} v2.0 \citep{drizzlepac} to remove distortion, reject cosmic rays, and construct coadded images. For the images discussed here, all drizzled products have a pixel scale of $0.03^{\prime\prime}$.

\begin{figure}[]
\begin{center}
\includegraphics[width=1\columnwidth]{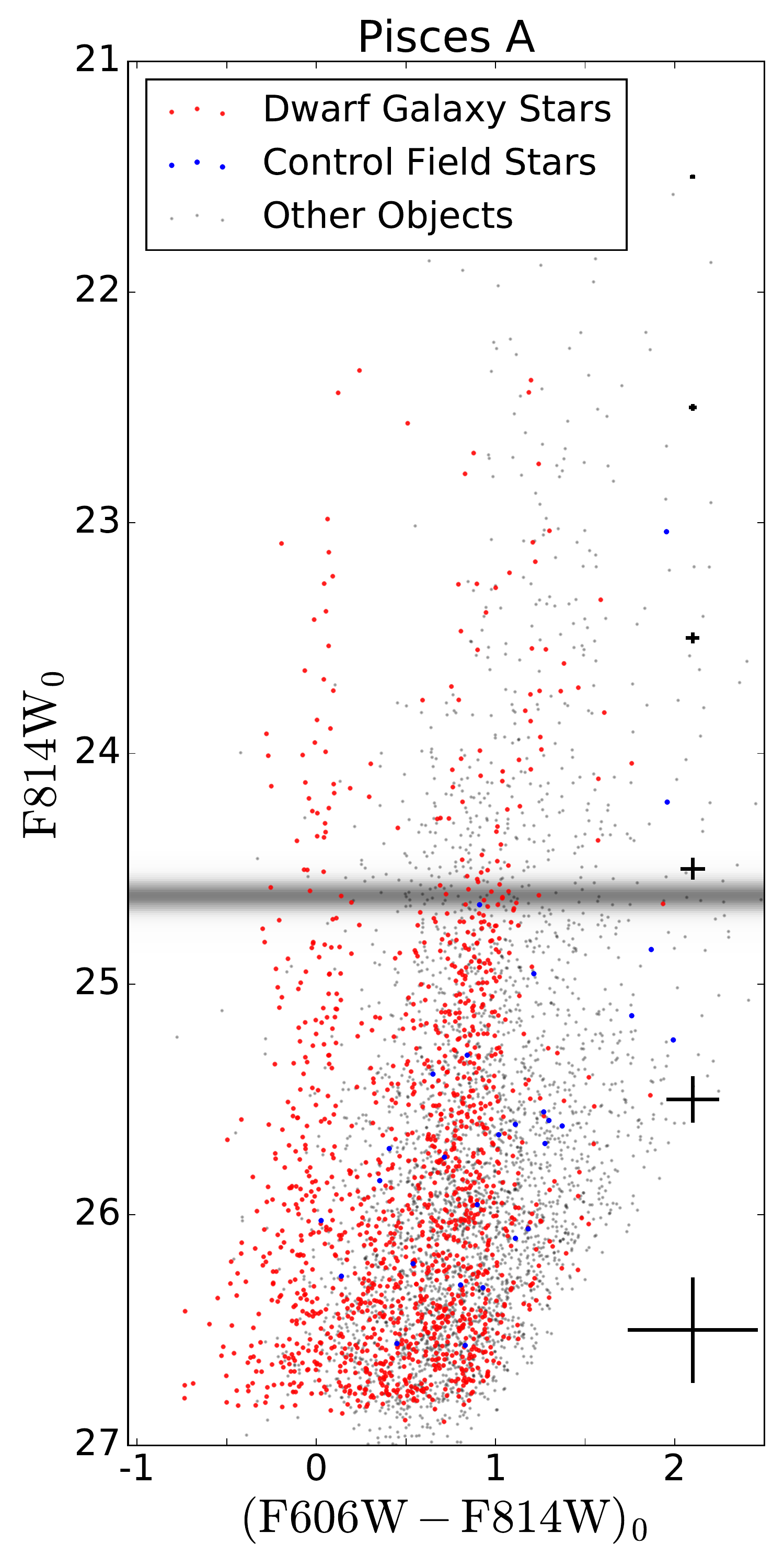}
\caption{ \protect\label{fig:CMDA}
F606W/F814W \ac{CMD} for the dwarf galaxy Pisces A.  Magnitudes are corrected for foreground extinction as described in the text.  Only objects with star-like morphologies are shown (i.e., \texttt{DOLPHOT} object type=1 or 2).  Larger red circles are stars within 25 arcsec (500 \ac{ACS}/\ac{WFC} pixels) of the center of Pisces A, blue circles are for an equal-size control field (on the other \ac{WFC} chip), while black points are the remainder of detected objects in the field.  Representative error bars for individual stars are shown on the right. The horizontal shaded band shows the probability distribution for the \ac{TRGB} location, derived as described in \S \ref{sec:dist}.}
\end{center}
\end{figure}

\begin{figure}[]
\begin{center}
\includegraphics[width=1\columnwidth]{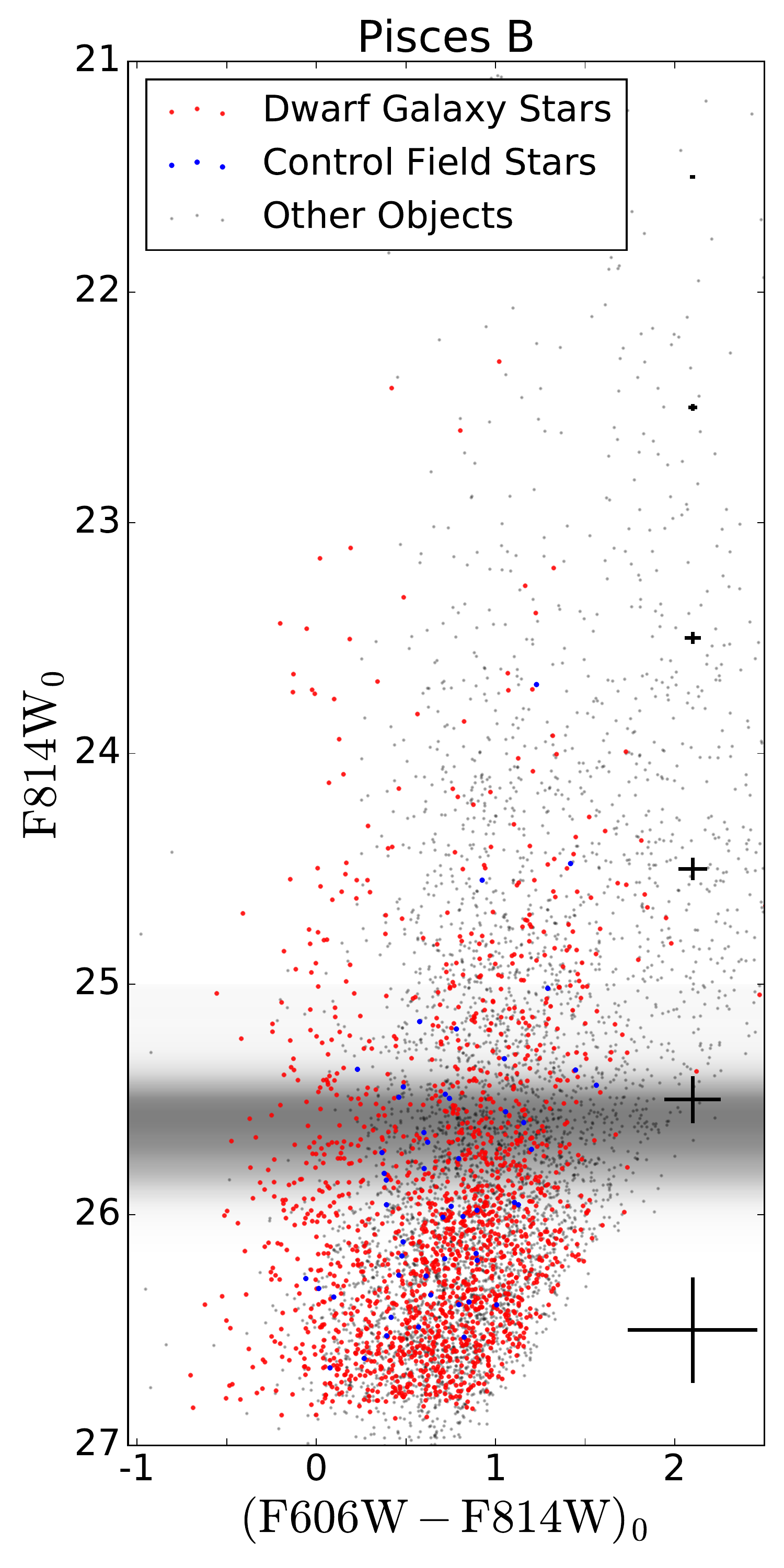}
\caption{ \protect\label{fig:CMDB}
Same as Figure \ref{fig:CMDA}, but for Pisces B.}
\end{center}
\end{figure}

We perform PSF-fitting photometry using \texttt{DOLPHOT} v2.0 (with the ACS module), a modified version of HSTphot \citep{hstphot}. We use the F606W/F814W bands directly for all analysis (we only convert to V and I for comparison with other datasets).  In what follows we only consider objects that have a \texttt{DOLPHOT} object type of 1 or 2 (using the default definitions for \texttt{DOLPHOT/ACS}), and are thus at least consistent with the \ac{PSF}.  While a small number of these objects may be non-stellar objects like compact background galaxies, in practice the vast majority will be member stars due to the compactness of these dwarf galaxies. For visualization and structural analysis, we correct these magnitudes for foreground extinction and reddening using the \citet{sf11} calibration of the \citet{sfd98} dust maps.  We do not use any objects that are in the chip gap for any exposure (although this would not affect our results because both target galaxies are far from the chip gap). Where relevant, we estimate photometric uncertainties using \acp{AST} in a \ac{CMD} box covering the \ac{RGB} and blue plume features for our galaxies. The resulting \acp{CMD} for Pisces A and B are shown in Figures \ref{fig:CMDA} and \ref{fig:CMDB}, respectively.

It is immediately apparent from these images and the \acp{CMD} that Pisces A and B are relatively nearby star-forming dwarf galaxies, confirming the result of \citet{T15}.  They have well-resolved stars, both show \acp{RGB}, blue plumes, irregular morphologies, and possible \ion{H}{2} regions with young star clusters.  To determine \emph{physical} parameters for these galaxies, however, an accurate distance measure is required.

\section{\ac{TRGB} Distances}
\label{sec:dist}

To determine distances, we make use of the \ac{TRGB} technique. The peak luminosity of the \ac{RGB} is a standardizable candle because it is driven by core helium ignition, so it provides a useful distance estimate for galaxies with an old stellar component that are close enough that the \ac{RGB} can be resolved \citep[e.g.,][]{ibenrenzini83,mcconn04}.  Here we describe our method for modeling the \ac{TRGB} and deriving a distance, and apply this method to our photometry of Pisces A and B.  Note that the code we use to infer probability distributions for the \ac{TRGB} of these galaxies is general enough to be applied to similar datasets, and we therefore make it publicly available at \url{https://github.com/eteq/rgbmcmr}.

\subsection{TRGB Model and Fitting Approach}
\label{ssec:distmethod}

\begin{figure}[]
\begin{center}
\includegraphics[width=1\columnwidth]{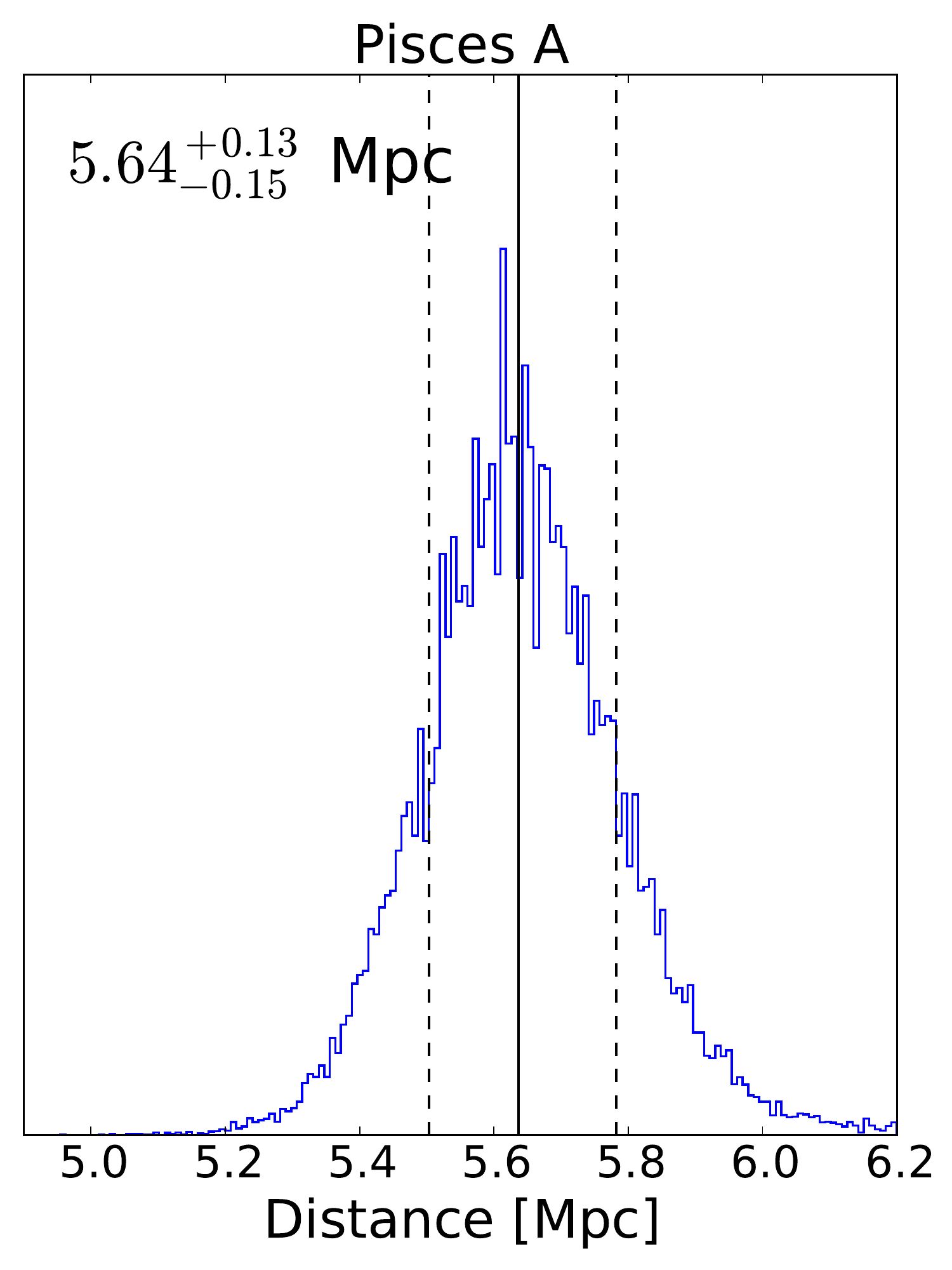}
\caption{ \protect\label{fig:distA}
Posterior probability distribution for the line-of-sight distance to Pisces A. The vertical solid line is the median, while the vertical dashed lines give the 84th/16th percentiles of the distribution.}
\end{center}
\end{figure}

\begin{figure}[]
\begin{center}
\includegraphics[width=1\columnwidth]{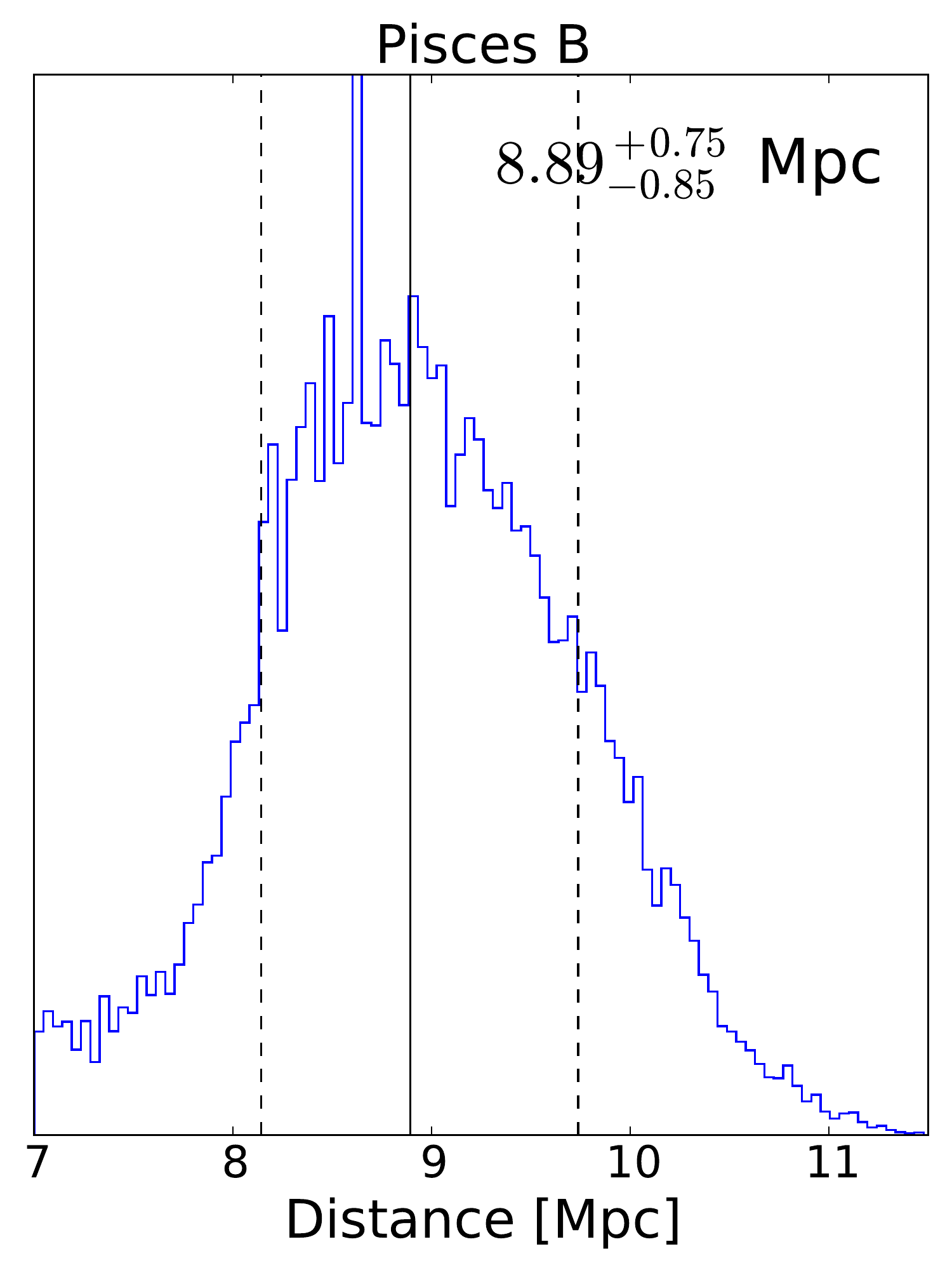}
\caption{ \protect\label{fig:distB}
Same as Figure \ref{fig:distA}, but for Pisces B.}
\end{center}
\end{figure}

Recent literature describes a variety of approaches for determining the magnitude of the \ac{TRGB} \citep[see][and references therein]{makarov06}.  Motivated in part by the work of \citet{conn11}, here we describe an approach using Bayesian inference on the photometric data described in \S \ref{sec:obs} to constrain the magnitude of the \ac{TRGB} and therefore the distance to our targets.  Such an approach is of particular importance for this paper because the posterior probability on the distances turn out to be non-Gaussian, particularly for Pisces B (see Figures \ref{fig:distA} and \ref{fig:distB}).  Hence, common approaches assuming Gaussian statistics are suspect and may be invalid.

Our assumed model for the \ac{TRGB} is motivated by the \ac{RGB} luminosity function parameterization of \citet{makarov06}. They adopt a model composed of a broken power law (in luminosity) with an offset at the slope discontinuity, smoothed by the photometric uncertainty and completeness corrections determined by \acp{AST}.  They also demonstrate that this model (with a maximum likelihood fitting technique) recovers the \ac{TRGB} for a variety of different stellar populations and foregrounds. We therefore adopt this model as a parameterization of the luminosity function near the \ac{TRGB}. Explicitly, the per-star likelihood function is,

\begin{multline}
\mathcal{L}_{\rm RGB}(m_i \mid \mu, \alpha, \beta, f) = \\
N \int_{m_1}^{m_2} \phi_{\rm RGB}(m, \mu, \alpha, \beta, f) \mathcal{N}[m_i - m + b(m), \sigma(m)] \mathcal{C}(m) dm \
\label{eqn:rgbmodel}
\end{multline}

\noindent where the luminosity function $\phi_{\rm RGB}$ is,

\begin{multline}
\phi_{\rm RGB}(m, \mu, \alpha, \beta, f) \equiv \\
     \begin{cases}
      \frac{ \alpha e^{\alpha(m-\mu)}}{e^{\alpha (m_2- \mu)}-e^{\alpha (m_1- \mu)}}  & m > \mu  \\
      f  \frac{\beta e^{\beta(m-\mu)}}{e^{\beta (m_2- \mu)}-e^{\beta (m_1- \mu)}}  & m \leq \mu \
     \end{cases}
\label{eqn:lumfunc}
\end{multline}

In these expressions, $m_i$ are the per-star extinction-corrected magnitudes (i.e., the data), $m_1$ and $m_2$ are the limits of the data, $N$ is a normalization constant (determined by numerical integration), $\mu$ is the magnitude of the \ac{TRGB}, $\alpha$ is the power law slope for the \ac{RGB} component, $\beta$ is the power-law slope for the non-\ac{RGB} population, typically \ac{AGB} stars, $f$ is the fraction of stars in the non-\ac{RGB} relative to those in the \ac{RGB}, and $\mathcal{N}$ is the standard Gaussian distribution. Also appearing in these expressions are the \ac{AST}-derived functions $b(m)$, $\sigma(m)$, and $\mathcal{C}(m)$. The first two are the first and second moments of the \ac{AST} magnitude distributions, given an input magnitude $m$, where the first is determined by a spline fit to the average bias, and the latter by a fit to $1.458$ times the median absolute deviation, a robust estimator for the standard Gaussian $\sigma$ \citep{madpaper}. $\mathcal{C}(m)$ is the magnitude-dependent completeness function from the \acp{AST}, parameterized by a sigmoid function.

The only parameter of interest in this model for determining the distance is the (unobserved) $\mu$, so we use Bayesian inference with this model to determine posterior probabilities for $\mu$.  We take Equation \ref{eqn:rgbmodel} as our per-star likelihood function.  We use uniform priors on $f$ of $\unifdistsymb(0,1)$, and for $\beta$ we use $\unifdistsymb(0, 3)$.  For $\mu$ we adopt slightly more informative priors of $\unifdistsymb(23.5, 25)$ and $\unifdistsymb(25, 26.5)$ for Pisces A and B, respectively.  This is justified by our prior knowledge that \ac{RGB} features are expected for a galaxy with an old component to its stellar population, and the features visible in these magnitude ranges of Figures \ref{fig:CMDA} and \ref{fig:CMDB} are typical of \acp{RGB}. For $\alpha$, we use the uninformative $\unifdistsymb(0, 3)$ for Pisces A.  However, for Pisces B, experimentation demonstrated that such a wide prior led to multiple modes in the posterior with apparently unphysical distributions (i.e., distributions with $f$ near $0$ or $1$, or $\alpha$ pinned to $0$).  Hence, for $\alpha$ of Pisces B, we adopt a Gaussian prior of $\mathcal{N}(0.3 \ln(10), 0.25)$.  This is motivated by previous studies which found that slopes of $\sim 0.3$ for a base-10 power law luminosity function often provide a decent fit to the \ac{RGB} \citep{mendez02,makarov06}.

The data in the Equation \ref{eqn:rgbmodel} likelihood are the observed \ac{RGB} star magnitudes.  For this work, we use the F814W magnitudes in a $(F606W-F814W)_0$ region of $(0.4, 1.5)$ for Pisces A and $(0.6, 1.1)$ for Pisces B.  We model extinction for these magnitudes using the \citet{sf11} calibration of the \citet{sfd98} dust maps.  We also apply magnitude limits of $F814W<26$ and  $F814W<26.5$ for Pisces A and B, which serves primarily to remove stars faint enough that completeness is a serious concern while still keeping enough stars to sample the tip reasonably. For a given set of parameters we numerically compute the integral of Equation \ref{eqn:rgbmodel} over a grid of $m$ using the trapezoid rule, and take the likelihood as the product of this over all the stars. For better numerical stability this is implemented as the sum of the natural log of the likelihoods.

With these pieces in place, we use the affine-invariant \ac{MCMC} sampler {\it emcee} \citep{emcee} to sample from the posterior distribution of the model parameters (i.e., the product of the likelihood and the parameter priors). We initialize the ensemble of walkers by sampling from the prior distributions.  Then for both galaxies we burn-in the sampler for 1500 samples, ensuring $>30$ autocorrelation times.  We obtain 5000 samples after burn-in to provide the samples from the posterior used in this paper.

This approach provides samples from the posterior distribution of $\mu$.  As a consistency check, we directly compare this to an approach like that of \citet{makarov06} and find that both methods provide $\mu$ values consistent at the $68 \%$ confidence/credible interval.  However, our quantity of \emph{physical} interest is the line-of-sight distance to the dwarf galaxies.  Hence we require a calibration of the  \ac{TRGB} to go from magnitude to distance. We adopt the \citet{rizzi07} calibration, as it provides calibrations in native \ac{ACS} magnitudes based on data similar to that used here.  While the intrinsic luminosity  of the \ac{TRGB} is relatively insensitive to stellar populations, there is some dependence on metallicity, which \citet{rizzi07} accounts for through a correction to the \ac{TRGB} magnitude that is color-dependent.  Hence, to determine a distance, we also need an estimate of the color of the \ac{TRGB}.

At first it might seem natural to build a color into the model in Equation \ref{eqn:rgbmodel}.
However, this would require a much more complicated (and not well-motivated) model of the shape of the \ac{RGB} because we would need to model the full \ac{CMD} rather than just a luminosity function.
Hence, we use a simple estimate for the color of the \ac{RGB}: for a particular value of $\mu$, we simply take the median $(F606W-F814W)_0$ value of the brightest $25$ \ac{RGB} stars for a given assumed $\mu$.
This provides a direct mapping from $\mu$ to $(F606W-F814W)_0$ color, allowing us to use the \citet{rizzi07} calibration to determine a distance to the galaxy.
For each sample of the $\mu$ posterior, we do this calculation, and treat the resulting distribution as a posterior on the distance to the galaxy.

\subsection{Distances to Pisces A and B}

In Figures \ref{fig:distA} and  \ref{fig:distB}, we show the posterior distributions on the line-of-sight distance to Pisces A and B, inferred using the approach described in \S \ref{ssec:distmethod}, as well as their median and ``$1\sigma$'' tails. It is immediately apparent that, while Pisces A has a distribution that is approximately Gaussian, the Pisces B distribution is not, with both skew and a flatter peak. This in part justifies our inferential approach, as it means any use of the distance to Pisces B and its uncertainties is most easily accomplished by using these samples directly.

\begin{figure*}[]
\begin{center}
\includegraphics[width=2\columnwidth]{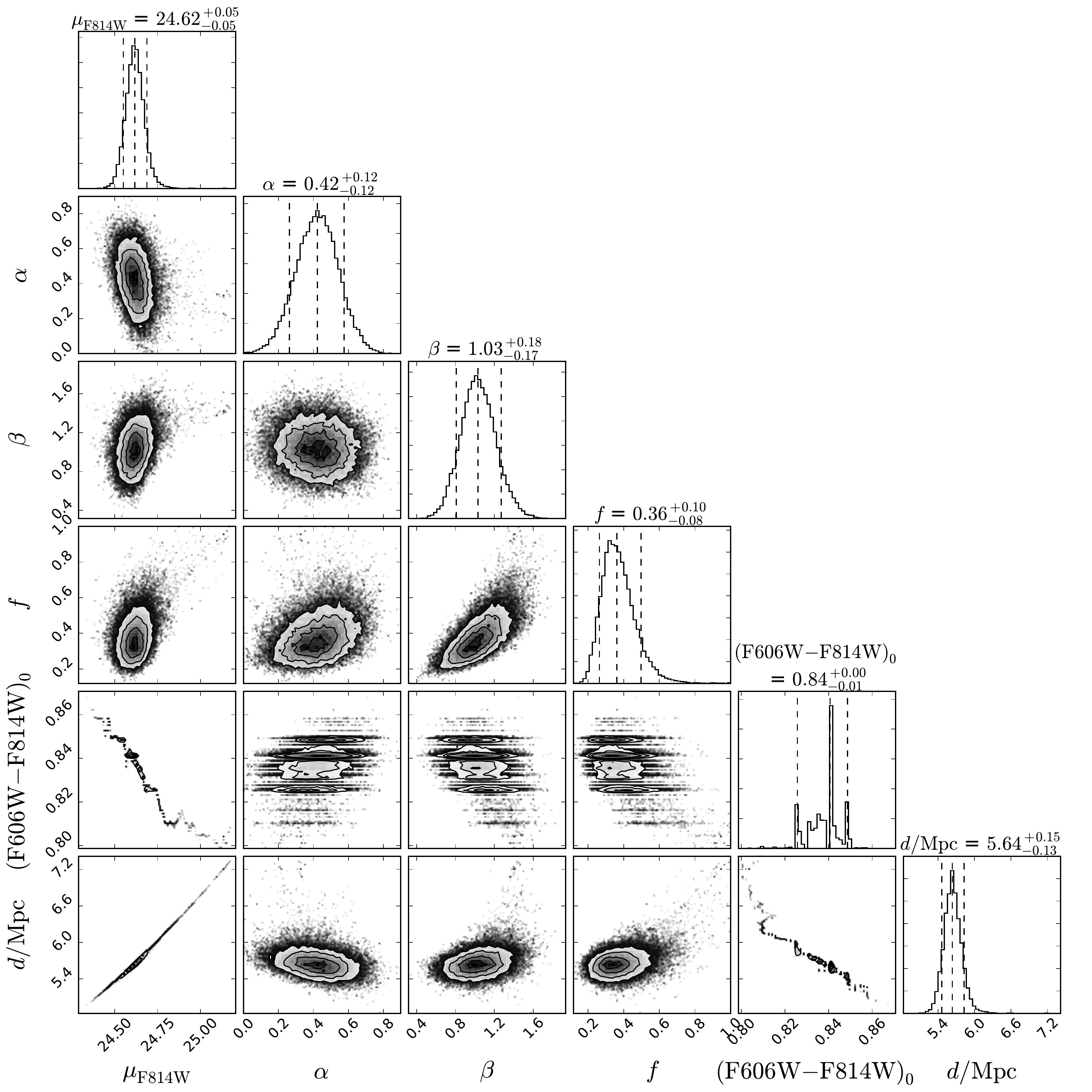}
\caption{ \protect\label{fig:postcovA}
Samples from the posterior distributions for the parameters of the TRGB model, constrained by the Pisces A dataset (see description of model in \S \ref{ssec:distmethod}).}
\end{center}
\end{figure*}

\begin{figure*}[]
\begin{center}
\includegraphics[width=2\columnwidth]{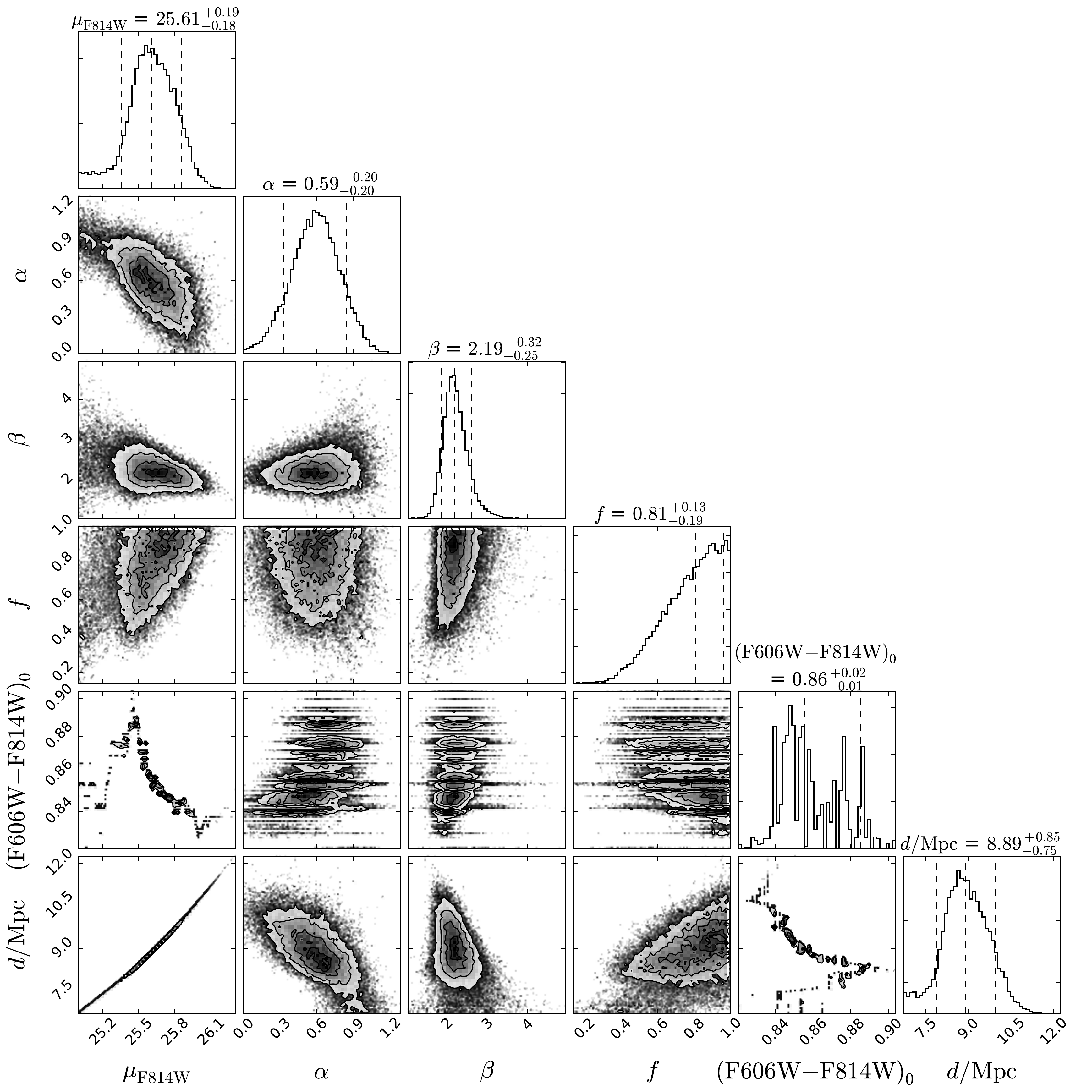}
\caption{ \protect\label{fig:postcovB}
Same as Figure \ref{fig:postcovA}, but for Pisces B.}
\end{center}
\end{figure*}

We show the covariances between the parameters of our model (Equation \ref{eqn:rgbmodel}) in Figures \ref{fig:postcovA} and \ref{fig:postcovB}.
The strongest features here are the discrete ``banding'' of the covariances with color of the TRBG, as well as the strong correlation between distance and the color.
These are a direct result of our simplistic model for the color, which is simply the median of the 25 stars nearest the tip.
The ``banding'' in the color distributions is due to discrete jumps when $\mu$ crosses below the magnitude of a bright star: the star is then removed form the list of 25 and a fainter star is added below.
This also explains the strong correlation between color and $\mu$: the latter deterministically sets the former as the set of 25 stars are medianed to yield the color.
The absence of these features in the distance distribution supports the claim in \S \ref{ssec:distmethod} that the color correction is sub-dominant over the other sources of uncertainty for these objects.  Another useful result apparent in Figure \ref{fig:postcovB} is an explanation for the tail to lower distances in the Pisces B distance distribution.  The low distances are strongly correlated with high values of $\alpha$.  Applying a stronger prior on $\alpha$ would remove this tail, as has implicitly been done in studies that assume $\alpha \sim 0.7$ \citep[$a \sim 0.3$ in base-10 power laws, e.g.,][]{mendez02}. We choose not to apply such a prior here, as we believe it is a \emph{possible} real feature.  That is, an alternative interpretation is that Pisces B has an anomalously high \ac{RGB} slope, rather than a detection of both \ac{RGB} and \ac{AGB}. Hence, we allow the posterior distance distribution to include this possibility, although with a low probability, as dictated by the data.

Regardless of the details, both distance measurements are quite peaked, implying a detection and providing reasonable distance estimates from which to infer physical parameters.  Pisces A is at $\sim 5.6$ Mpc, and Pisces B $\sim 9$ Mpc. These distances are comparable to the upper ``Hubble Flow'' estimates from \citet{T15}, although slightly more distant for Pisces A, most likely due to either peculiar motions or deviations of the local Hubble Flow due to the \ac{LG} \citep{pen14}.  The distance posteriors yield very high ($\gtrsim 10^5:1$) odds that both galaxies are in the Local Volume ($2 < d < 12$ Mpc).

\section{Structural Parameters}

\label{sec:sparams}

While \citet{T15} determined structural parameters for Pisces A and B from the \ac{SDSS}, the  higher resolution and deeper HST imaging provides an opportunity for substantial improvement.
Furthermore, these observed parameters can be converted to much improved physical parameters using the distance determination of \S \ref{sec:dist}.
In this section, we perform this exercise, first describing an approach to determine observed structural parameters using a  Bayesian inference, and then presenting the resulting parameter distributions.

To make full use of the (non-Gaussian) distance distributions from \S \ref{sec:dist}, we adopt a similar inference approach to determine the as-observed structural parameters.
It may at first seem appropriate to infer structural parameters from the same catalogs we use in \S \ref{sec:dist}.
However, subtracting all the point sources in the catalogs reveal an unresolved component in both Pisces A and B.
If the \ac{SFH} of these galaxies is not uniform (as demonstrated in \S \ref{sec:sfhs}), the bright resolved stars may have a different structural parameter distribution than the unresolved component.
Hence, we fit our model directly to the drizzled per-pixel flux data, rather than the photometric catalogs.

The particular model we adopt here is an elliptical {S{\'e}rsic} profile \citep{sersic} with the {S{\'e}rsic} index left as a free parameter.
This model is flexible enough to accomodate a variety of profiles that are good fits for some of the faintest \ac{LG} galaxies \citep[e.g.][]{martin08, munoz12, mcconlgcat}, while providing easy comparison to larger surveys where integrated light measurements often assume exponential or de Vaucouleurs profiles \citep[e.g.,][]{dr7}.
The parameterization we use has seven parameters: the total flux $F$, the {S{\'e}rsic} index $n$, the major-axis effective radius
$R_{\rm eff,maj}$ (i.e., the radius encompassing half the light if $e=0$), the coordinates of the center of the profile $x_0$ and $y_0$, the ellipticity ($e=1-b/a$), and the position angle of the ellipse $\theta$.

{\it AstroDrizzle} provides an option (the {\it ERR} weight type) to output per-pixel uncertainties.
This takes advantage of the long experience with \ac{ACS} to account for the various contributions from poisson noise, read noise, dark current, sky, etc.
Hence, we assume each pixel is Gaussian distributed about the model with inverse variance set by the drizzled weight map.
Combined with our model, the per-pixel likelihood is,

\begin{multline}
\mathcal{L}_{\rm phot}(I_i, \sigma_{I,i} \mid F, n, R_{\rm eff,maj}, x_0, y_0, e, \theta) = \\
     \mathcal{N}(I_i-S(F, n, R_{\rm eff,maj}, x_0, y_0, e, \theta), \sigma_{I,i})
\label{eqn:strucparammodel}
\end{multline}

where $I$ is the per-pixel count rate and $S$ is the {S{\'e}rsic} profile discussed above.

We then take as the likelihood function the product over all pixels in a window much larger than the visible extent of the galaxies for each of the 4 drizzled images (F606W and F814W for both Pisces A and B).
As in \S \ref{sec:dist}, we use {\it emcee} to sample from the posterior distributions of each parameter in this likelihood given an assumed set of priors.

For all four images, we use uniform priors on all parameters other than flux.
We adopt an $R_{\rm eff,maj}$ prior of $\unifdistsymb(150, 800)$ (pixels), a prior on $e$ of $\unifdistsymb(0.25, 0.75)$, and a $\theta$ prior of $\unifdistsymb(90, 180)$.
For $x_0$ and $y_0$ we visually identify an approximate center and then choose uniform priors ranging from 50 pixels on either side of the visually identified point.
For $n$, experimentation revealed that a wide prior like $\unifdistsymb(0.2, 5)$ leads to multimodal posteriors where the high-$n$ solution was clearly incorrect, typically centered on a bright star or similar feature much more compact than the galaxies.
Hence, we adopt the slightly more restrictive prior $\unifdistsymb(0.2, 1.5)$.
For $F$, we adopt a Gaussian prior centered on a total flux in a circular aperture chosen by-eye to cover the unresolved light in the \ac{PSF}-subtracted images, but wide enough to make the prior only weakly informative.
Specifically, for Pisces A we use $\mathcal{N}(4000, 800)$ for the F606W image and $\mathcal{N}(3700, 800)$ for F814W, while for Pisces B we use $\mathcal{N}(5800, 1000)$ for both.
In practice, all of these priors are relatively uninformative for all four images: the posteriors are much narrower than the input priors, implying they are data-driven and not prior-dominated.

We compare our derived fluxes to corresponding by-eye estimated (elliptical) aperture magnitudes and find that they are quite close to the quantitative measurement following the above technique.
To more quantitatively test the validity of this technique for semi-resolved galaxies, we perform a mock data experiment.
We assume a power law stellar luminosity function  with a slope roughly comparable to that observed for Pisces B.
Then we sample stars from an elliptical {S{\'e}rsic} profile, which yields an image with both resolved stars and an unresolved component (i.e., the sum of many stars that are individually below the noise floor).
From this ``perfect'' image, we randomly sample assuming a Poisson distribution with noise characteristics matched to the Pisces B \ac{ACS} observations.
We then run the structural parameter sampling procedure described above on this mock dataset.
This experiment yields final posteriors that are consistent with the input parameters (i.e. they all lie well within $1\sigma$ bounds), and have widths similar to those of the real observations.
See \url{https://github.com/eteq/piscdwarfs_hst/blob/master/Mock_Structural_Parameters.ipynb} for full details of this experiment.

\begin{figure*}[]
\begin{center}
\includegraphics[width=2\columnwidth]{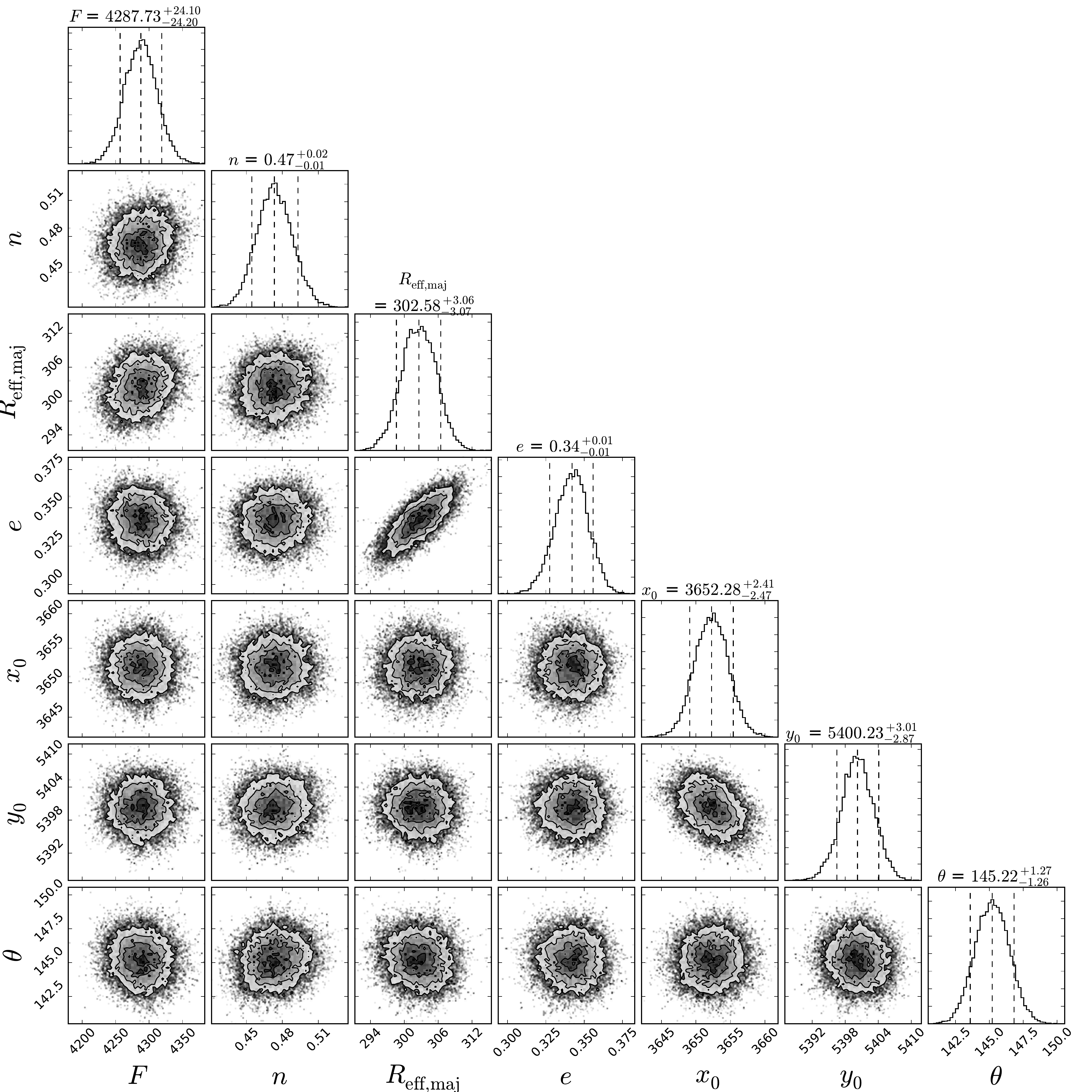}
\caption{ \protect\label{fig:structparamA}
Samples from the posterior distributions for the structural parameters model (see description of model in \S \ref{sec:sparams}) for the F606W image of Pisces A.  $R_{\rm eff, maj}$, $x_0$, and $y_0$ are in in pixels, which correspond to $0.03"$ for the drizzling parameters used here.  The total flux $F$ is in \ac{WFC} counts per second, and $\theta$ is in degrees. Also note that $\theta$ here is towards the pixel x-axis from the pixel y-axis, which is \emph{counter-clockwise} in Figures \ref{fig:piscAimg} and \ref{fig:piscBimg}.}
\end{center}
\end{figure*}

\begin{figure*}[]
\begin{center}
\includegraphics[width=2\columnwidth]{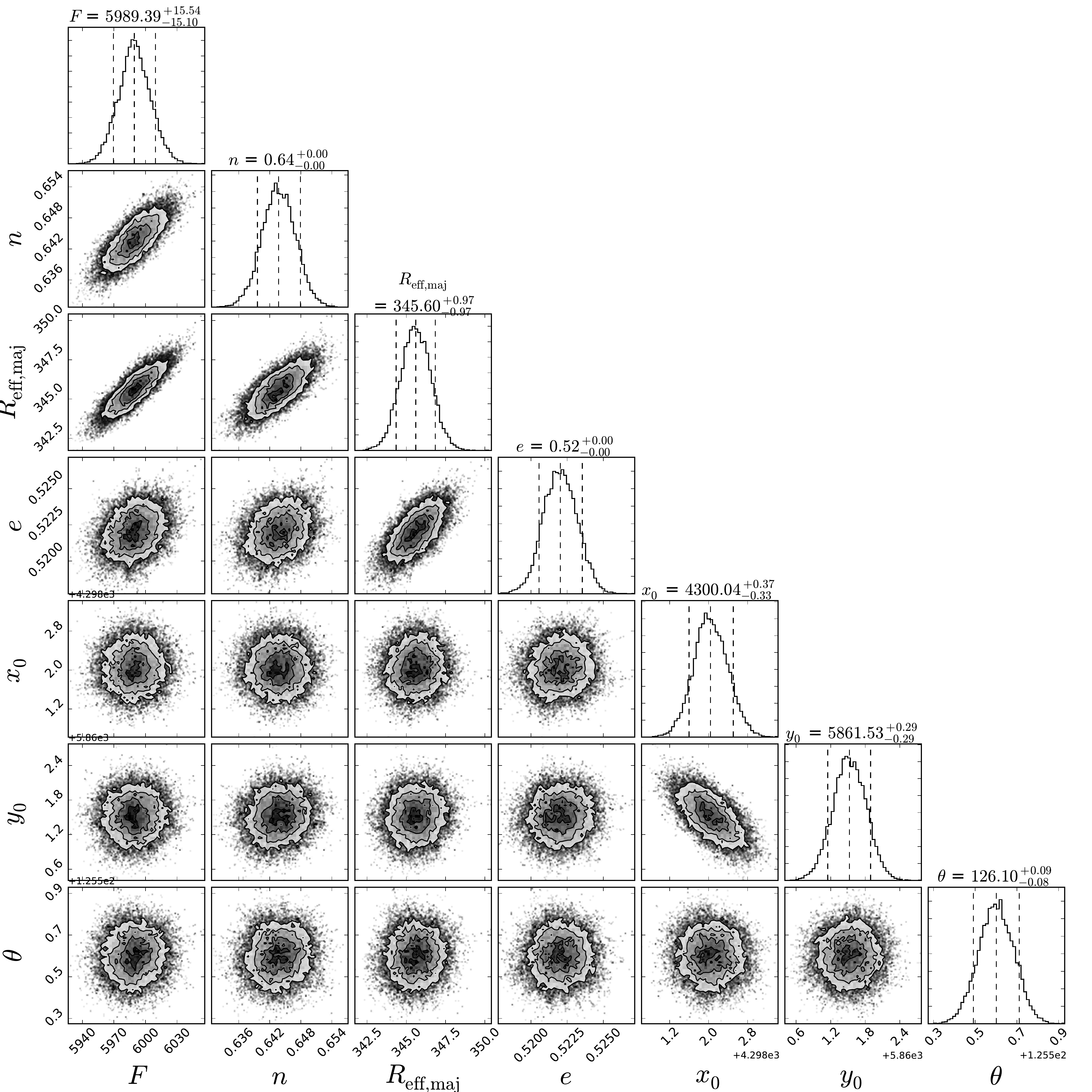}
\caption{ \protect\label{fig:structparamB}
Same as Figure \ref{fig:structparamA}, but for Pisces B.}
\end{center}
\end{figure*}

In Figures \ref{fig:structparamA} and \ref{fig:structparamB}, we show the sampled distributions and their covariances for the F606W images of Pisces A and B, respectively.
We do not show similar plots for F814W because their behavior is very similar and the actual values are provided in Tables \ref{tab:res} and \ref{tab:quantities}.
It is immediately clear from these figures that the parameters are well-constrained and quite Gaussian, although with some covariances.

The structural parameters derived in this manner are as-observed quantities, however, and our goal is physical parameters.
Two steps remain: calibrating the count rate to physical flux/apparent magnitude, and using the distances from \S \ref{sec:dist} to convert from observed to physical parameters.
The calibration step is straightforward: we simply adopt the \citet{acszpt} zero points provided on the ACS web site\footnote{\url{http://www.stsci.edu/hst/acs/analysis/zeropoints}}.
To convert to $V$ and $I$-band estimates (only for Table \ref{tab:res}), we use the transformations of \citet{acstrans05}, \S 8.3.
The final step to yield physical parameters requires combining the distance distributions shown as Figures \ref{fig:distA} and \ref{fig:distB} with the apparent magnitudes and $R_{\rm eff,maj}$ derived as described in this section.
We do this by simply randomly assigning a sample from the distance distributions to a sample from the structural parameters, and using that combination to derive the physical parameter.

After following the above procedure, the resulting distributions (with appropriate covariances) for absolute magnitude, colors, and $r_{\rm eff,maj}/{\rm pc}$ are provided as columns in Table \ref{tab:quantities}, and are summarized in Table \ref{tab:res}.
The $r_{\rm eff}$ for Pisces A is rather close to the ``Hubble Flow'' scenario of \citet{T15}, while for Pisces B it is $\sim 25\%$ larger.
For Pisces A, this is due to the competing effects of a larger distance but a smaller $R_{\rm eff}$, while for Pisces B the primary effect is a larger $R_{\rm eff}$.
The $R_{\rm eff}$ changes are likely due to the combined effect of much lower backgrounds and less confusion in the \ac{HST} images.
In contrast, the absolute magnitudes are comparable to the estimates from \citet{T15} for the ``Hubble Flow'' scenario, although in this work, Pisces A is somewhat more luminous due to the larger distance compared to \citet{T15}.

\section{Star Formation Histories}
\label{sec:sfhs}

\begin{figure}[]
\begin{center}
\includegraphics[width=1\columnwidth]{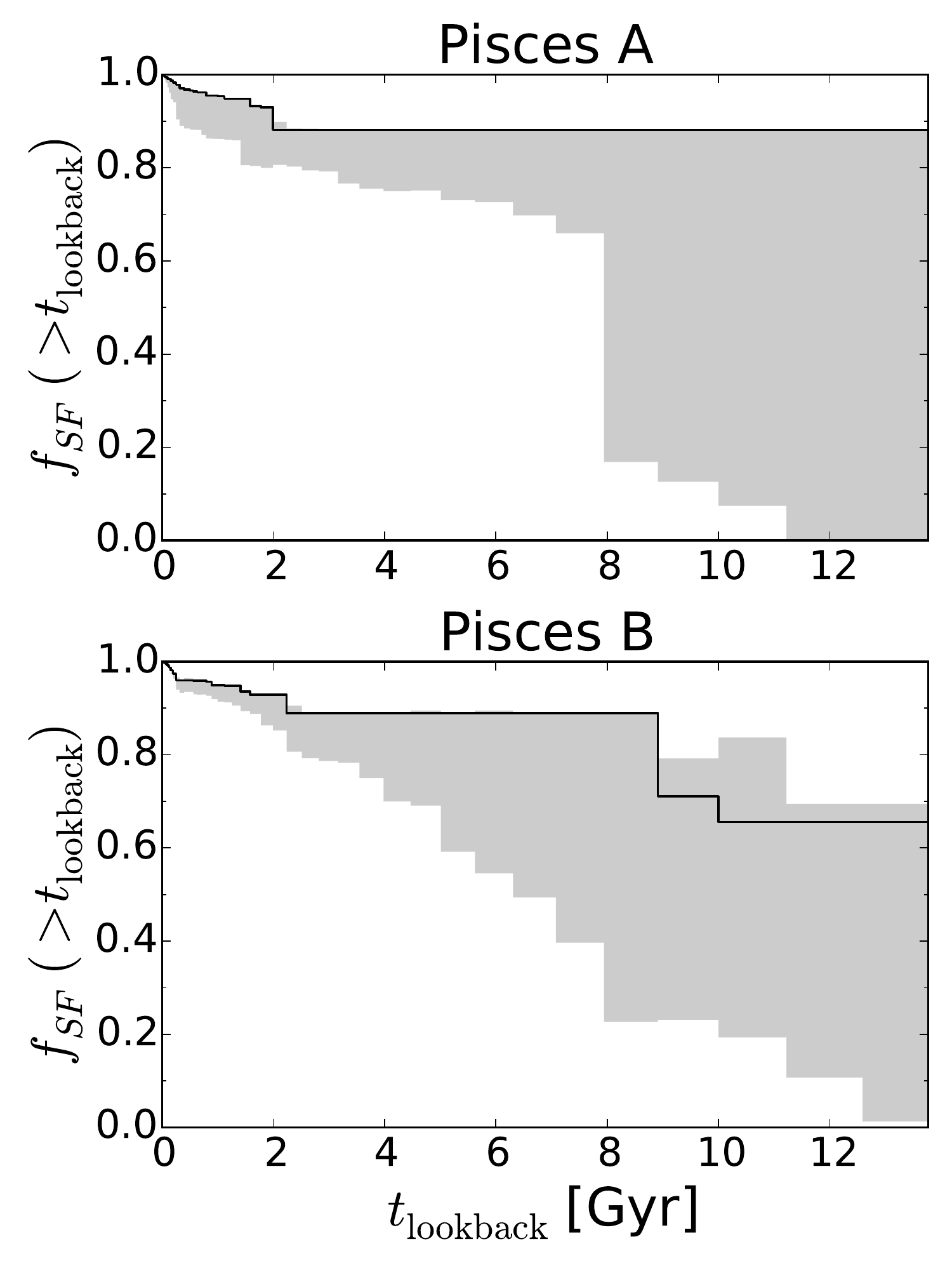}
\caption{ \protect\label{fig:sfhs}
Cumulative (fractional) star formation histories for Pisces A and B, derived from the approach described in \S \ref{sec:sfhs}.
Note the increase at late times ($<300 \, \rm{Myr}$ ago).  The black line is the most probable history, while the shaded region are 16/84th percentiles for the combination of random and systematic uncertainties.}
\end{center}
\end{figure}

To estimate \acp{SFH} of Pisces A and B, we use the approach described in \citet{weisz14}, applied to the \ac{ACS} photometry described in Section \ref{sec:obs}.
This approach uses the \ac{CMD} fitting routine \texttt{MATCH} \citep{dolphin02match} to infer the \ac{SFH} of a galaxy from resolved star photometry.
To estimate random uncertainties, we use confidence intervals from the \ac{MCMC} approach described in \citet{dolphin13}.
We also compute the systematics following the methods outlined in \citet{weisz11} and \citet{dolphin12}.
For more details on this procedure, see \citet{weisz14}.
While Figure \ref{fig:sfhs} shows the normalized star formation history for clarity, we also provide the absolute stellar mass inferred from this procedure for each galaxy in Table \ref{tab:res}.

In Figure \ref{fig:sfhs}, we show the intervals for the cumulative star formation in Pisces A and B, including both the random and systematic uncertainties.
There is large uncertainty, particularly for lookback times $>3$ Gyr, because the \acp{CMD} are relatively shallow.
However, both galaxies are consistent with a recent ($\lesssim 300$ Myr) increase in star formation.
While this is not surprising at a qualitative level given their HI content and blue colors, it is borne out here quantitatively.

\begin{figure*}[]
\begin{center}
\includegraphics[width=2\columnwidth]{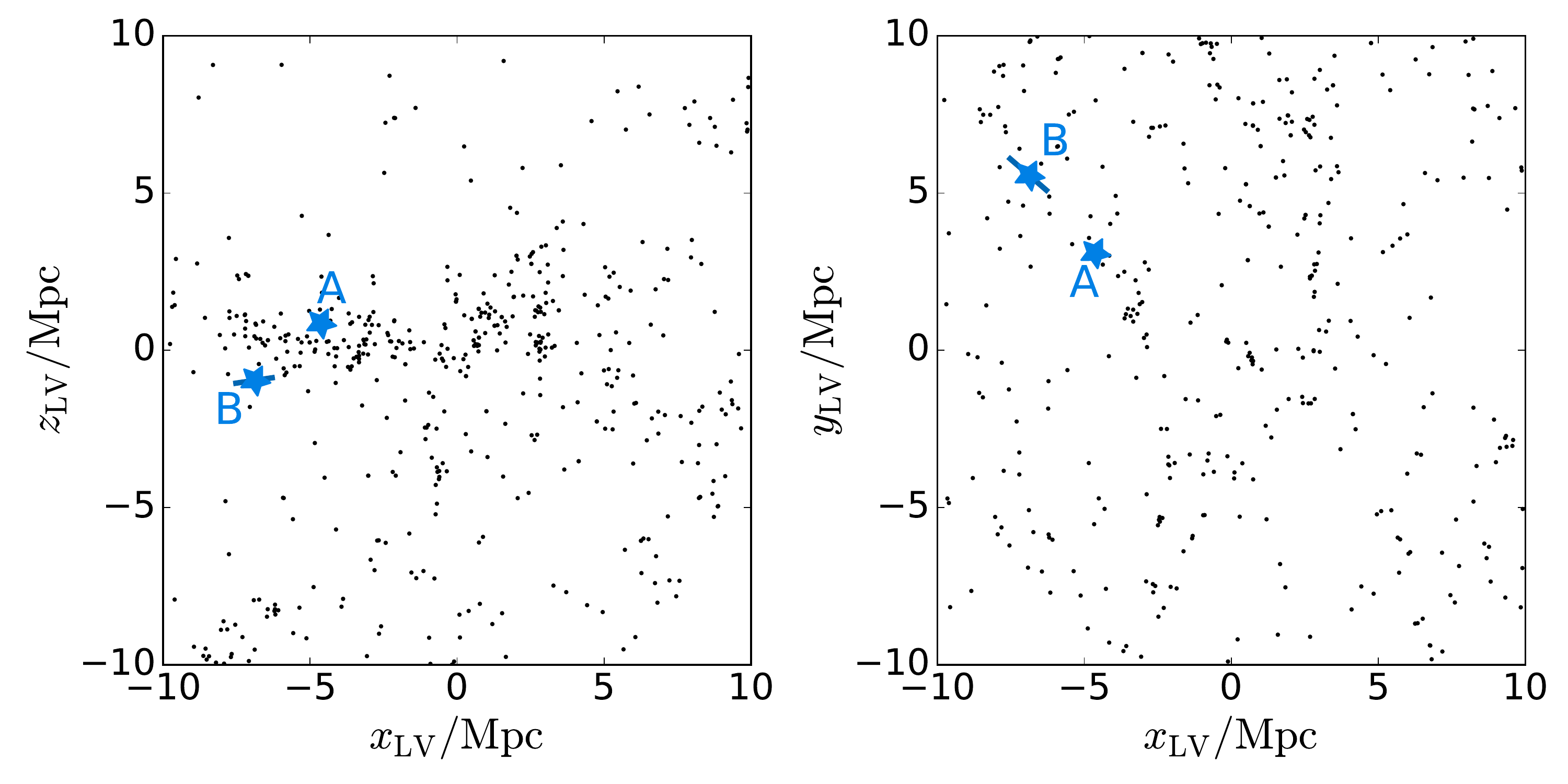}
\caption{ \protect\label{fig:position}
Location of Pisces A and B in the Local Volume.  Pisces A and B are the (blue) stars-shaped data points, with distance error bars, while the other points are a 10 Mpc volume-limited sample of nearby galaxies (see text). The coordinate system is oriented such that the center is at the Earth, and the z-axis points toward the Local Void \citep{localvoid}.}
\end{center}
\end{figure*}

\section{Discussion}
\label{sec:disc}

With complete posteriors on the distance to Pisces A and B (\S \ref{sec:dist}), their structural parameters (\S \ref{sec:sparams}), and  estimates of their \acp{SFH} (\S \ref{sec:sfhs}), we are now in a position to provide the context in which these galaxies evolved.
In Figure \ref{fig:position}, we show a 3D cartesian representation of galaxies in the Local Volume.
The underlying data (black points) are galaxies from a catalog comprised of the NASA-Sloan Atlas \citep{nsa}, the Extragalactic Distance Database \citep{EDD}, DR3 of the 6dF Galaxy Survey \citep{6df1, 6df2} and the 2MASS Extended Source Catalog \citep[XSC, ][]{2mass}.
These data sets have been combined and volume-limited to the detection limit of the 2MASS XSC at 10 Mpc ($M_K<-17.3$).
Pisces A and B are shown as blue star-shaped symbols with error bars showing distance uncertainties (84th/16th percentile).
The z-axis (shown in the left panel) is oriented to point in the direction of the Local Void \citep[$18^{\rm h}38^{\rm m} \; 18^{\circ}$, ][]{localvoid}, an orientation that highlights the fact that these galaxies both appear to be near the boundary of local filamentary structure. 
This is borne out more quantitatively by considering $d_5$, the distance to the 5th nearest neighbor in the catalog shown in Figure \ref{fig:position}. 
Pisces A and B are $74$th and $54$th percentile, respectively, placing them near the middle of the environments available in the Local Volume (compared to e.g. the Milky Way or M31, which are at $\sim 5$th percentile).

The location of these galaxies near void boundaries raises the possibility that they have spent most of their time in the surrounding voids, and are only now falling into a higher-density environment.
Direct evidence for this would require a velocity measurement for these galaxies in a direction perpendicular to the filament.
However, the line-of-sight vector is nearly parallel to the filament axis, so the line-of-sight velocity cannot be used to test if these galaxies are infalling.
Furthermore, proper motion measurements for objects this distant are not likely to be possible with current observational facilities.
Nonetheless, as discussed in \S \ref{sec:sfhs}, recent star formation of these galaxies may be a critical clue: they could potentially be triggered by infall onto the cosmic web \citep[e.g.,][]{webstripping13} or other interactions if they are only now entering a higher-density environment.
Furthermore, their HI properties are broadly consistent with typical void galaxies from \citealt{kreckel12} (although the galaxies in that sample are higher mass due to selection effects).
This along with the spatial orientation relative to the filament provide at least circumstantial evidence that these galaxies are both void galaxies only now entering a higher-density environment.

\begin{figure*}[]
\begin{center}
\includegraphics[width=1.5\columnwidth]{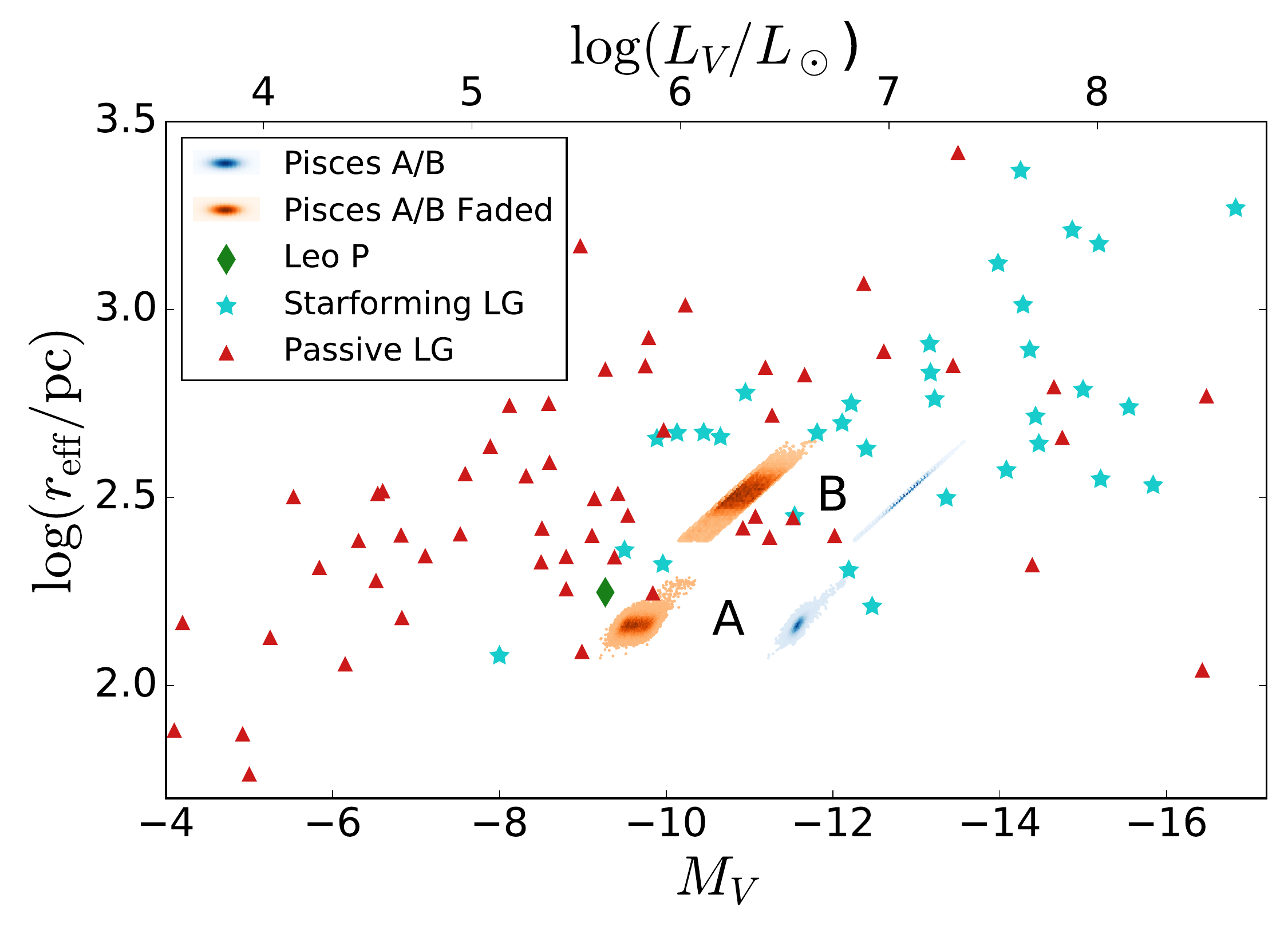}
\caption{ \protect\label{fig:context}
Size-luminosity relation for Pisces A and B, Leo P, and \ac{LG} galaxies.  The \ac{LG} galaxy sample is taken from the \citet{mcconlgcat} catalog, selected on gas content where available and otherwise color to determine if a galaxy is starforming (blue star-shaped symbols), or passive (red triangles).  The Leo P measurements (green diamond) are from \citet{mcquinn15}.  The blue shaded regions are probability distributions for Pisces A and B, including uncertainties from both the distance (\S \ref{sec:dist}) and structural parameter measurements (\S \ref{sec:sparams}).  The orange shaded regions are the same data, but faded assuming passive evolution, as they would appear in the future if quenched (see text).}
\end{center}
\end{figure*}

Another important contextual clue for these galaxies is provided in Figure \ref{fig:context}.
We show Pisces A and B in the size-luminosity relation (blue distributions), along with galaxies of the \ac{LG} (red triangles and cyan star-shaped symbols), and Leo P \citep[green diamon][]{leophst}.

While not quite as extreme as in Figure 3(a) of \citet{T15} due to the new distances, this figure shows that both are quite compact relative to typical starforming galaxies of the \ac{LG}.
This suggests the possibility that these galaxies are undergoing delayed evolution.
That is, if they are infalling from surrounding voids (as suggested by the above discussion), they spent most of their history in the void.
If this is so, the void environment would have slowed their evolution due to both lower gas densities and slower dynamical times \citep[e.g.,][]{aragoncalvo13}.
This is further supported by the fact that their HI content is somewhat high relative to similar galaxies\footnote{This statement comes from the results shown in \citet{T15} Figure 3b, but still holds for the new distances derived in this paper.}.

We also show a rough estimate of the size-luminosity relation expected if Pisces A and B were to passively fade (orange distributions).
That is, their morphology (here captured as $r_{\rm eff}$) and stellar masses were left unchanged, but the mass-to-light ratio are adjusted to be appropriate for an old stellar population.
While this as a rather extreme and unphysical model for stripping, a more detailed model is beyond the scope of this paper, and we use it here simply as an estimate of the magnitude of the effect.
This demonstrates that if Pisces A and B were allowed to fade through passive evolution, they would be more comparable to known passive galaxies of the \ac{LG}.
This then provides both an explanation for the anomalously compact sizes of these galaxies, and implies they are excellent candidates as ``prototypes'' of \ac{LG} galaxies: i.e., snapshots of what the present-day \ac{LG} dwarf galaxies may have been like at earlier epochs.

\section{Conclusions}
\label{sec:conc}

In this paper, we:

\begin{itemize}
  \item Describe observations of the Pisces A and B dwarf galaxies with the \ac{HST} \ac{ACS}, and clearly identify a resolved \acp{RGB}.
  \item Describe a Bayesian inferential method of determining distances to and structural parameters of Local Volume galaxies with resolved stellar populations, and apply this approach to our Pisces A and B data set. From this we infer the distance and photometric parameters of these galaxies along with the various parameter covariances.
  \item We conclude that Pisces A and B are Local Volume galaxies (at $5.6 \pm 0.2$ and $9.2 \pm 1.1$ Mpc, respectively).  Moreover, with the newly-constrained distances and an estimate of their \acp{SFH} derived from the same data, we find they are plausibly galaxies from the Local Void infalling onto filamentary structure in the Local Volume.
\end{itemize}

These galaxies (and others like them) thus represent a potentially valuable tool as ``initial conditions'' for dwarf galaxies of the \ac{LG} or other higher-density environments.

\vspace*{1.5 \baselineskip}

We acknowledge Frank van den Bosch and Anil Seth for helpful discussion about this work.
We also thank the anonymous referee for feedback that improved this paper.

This research made use of Astropy, a community-developed core Python package for Astronomy \citep{astropy}.  It also used the \ac{MCMC} fitting code {\it emcee} \citep{emcee}. It further made use of the open-source software tools Numpy, Scipy, Matplotlib, IPython, and corner.py \citep{numpyscipy, matplotlib, ipython, cornerpy}.
This research has made use of NASA's Astrophysics Data System.

Support for EJT and DRW was provided by NASA through Hubble Fellowship grants \#51316.01 and \#51331.01 awarded by the Space Telescope Science Institute, which is operated by the Association of Universities for Research in Astronomy, Inc., for NASA, under contract NAS 5-26555.

%
% \begin{deluxetable}{cc}
% %\tabletypesize{\footnotesize}
% \tablecolumns{2}
% \tablewidth{0pt}
% \tablecaption{ Distributions of Quantities \label{tab:quantities}}
% \tablehead{
% \colhead{Distance (A)} & $\mu$ (A) \\
% \colhead{Mpc} & \colhead{mag}
% }
% \startdata
% 1 & 2  \\
% ... & ... \\
% \enddata
% \vspace{-0.8cm}
% \tablecomments{Posteriors of quantities inferred as described in the text. The rows are aligned so as to preserve correlations when they are present between two quantities. This is a sample table to demonstrate the format.  The true table is available in the electronic edition of the journal. It will include the following quantities (for both Pisces A and B): }
% \end{deluxetable}

\begin{table}

\begin{center}
\caption{Distributions of Quantities}
\label{tab:quantities}

\begin{tabular}{c c c c}

  \hline
  \hline
Distance (A) & $\mu_{814}$ (A) & $({\rm F606W}-{\rm F814W})_0$ (A) & $r_{\rm eff}$ (A) \\
Mpc & mag & mag & kpc \\
  \hline
5.62 & 24.61  & 0.841 & 2.21\\
... & ... & ... & ... \\
  \hline

\end{tabular}
\end{center}

Posteriors of quantities inferred as described in the text. The rows are aligned so as to preserve correlations when they are present between two quantities. This is a sample table with a limited number of columns to demonstrate the format.  The true table is available in the electronic edition of the journal. It includes the following quantities (for both Pisces A and B):
Distance, F814W magnitude of the \ac{TRGB}, $({\rm F606W}-{\rm F814W})_0$ color of the \ac{TRGB}, $\alpha$, $\beta$, $f$ (from Equation \ref{eqn:rgbmodel}), F606W (total), F814W (total), S{\'e}rsic index (F606W),  ellipticity (F606W), position angle (F606W), $R_{\rm eff}$ (on-sky half-light radius in F606W),  $r_{\rm eff}$ (physical half-light radius in F606).
\end{table}

\bibliography{biblio}{}

\end{document}